\documentclass[a4paper, 10pt]{article}
\usepackage[T1]{fontenc}
\usepackage[utf8]{inputenc}
\usepackage{amsfonts}
\usepackage{amssymb}
\usepackage{indentfirst}
\usepackage{amsmath}
\usepackage{amsthm}
\usepackage {geometry}
\usepackage[pdftex]{graphicx}

\newgeometry{tmargin=4.65 cm, bmargin=4.65 cm, lmargin=3.65 cm, rmargin=3.75 cm}

\begin{document}
\setlength{\baselineskip}{13pt}
\newpage
\vfill \vfill \vfill
\begin{center}
\begin{footnotesize}
\textbf{GENERAL RELATIVITY WITH A POSITIVE COSMOLOGICAL CONSTANT $\Lambda $ AS A GAUGE THEORY}
\end{footnotesize}
\end{center}

\begin{center}
MARTA DUDEK\\
\begin{footnotesize}
\textit{Institute of Mathematics, University of Szczecin}\\
\textit{Wielkopolska 15, 70-451 Szczecin, Poland, EU}\\
\textit{e-mail: marta.dudek@vp.pl}
\end{footnotesize}
\end{center}

\begin{center}
JANUSZ GARECKI\\
\begin{footnotesize}
\textit{Institute of Mathematics and Cosmology Group, University of Szczecin}\\
\textit{Wielkopolska 15, 70-451 Szczecin, Poland, EU}\\
\textit{e-mail: garecki@wmf.univ.szczecin.pl}
\end{footnotesize}
\end{center}

\hspace{6 cm}
\\
\\
In this paper we show that the general relativity action (and Lagrangian) in recent Einstein-Palatini formulation is equivalent in four dimensions to the action (and Langrangian) of a gauge field. 

At first we briefly showcase the Einstein-Palatini (EP) action, then we present how Einstein fields equations can be derived from it. In the next section, we study Einstein-Palatini action integral for general relativity with a positive cosmological constant $\Lambda $ in terms of the corrected curvature $\Omega_{cor}$. We will see that in terms of $\Omega_{cor}$ this action takes the form typical for a gauge field. Finally, we give a geometrical interpretation of the corrected curvature $\Omega_{cor}$.

\vspace{0.5 cm}

\section{Introduction: Einstein-Palatini action for general relativity}

In this section we would like to remind you briefly about Einstein- Palatini formalism for general relativity (GR).

The Einstein field equations can be derived from postulation that the Einstein-Hilbert action is true action for GR. Albert Einstein firstly used only metric as an independent variable to do variation of this action. The connection with EH action, in this approach, is the metric and symmetric Levi-Civita connection. Later Einstein and Palatini proposed to take the metric and affine connection as independent variables in the action principle. This method allowed to compute the field equations for a more general metric affine connection rather than the Levi-Civita connection. Here the spacetime admitted torsion when matter Lagrangian explicite depended on connection.Thus, the Einstein-Palatini formalism gave us a powerful tool for theories of gravitation which have more general Riemann-Cartan geometry.\\
In this section, we are going to present Einstein-Palatini action in recent formulation and how the Einstein fields equations can be computed from it.\\
\\
The Einstein-Palatini action with cosmological constant $\Lambda $ in this new formulation [3] is defined as follows
\begin{align}
S_{EP}=\frac{1}{4\kappa}\int \limits_{\mathcal{D} }\Big( \vartheta^{i}\wedge \vartheta^{j}\wedge \Omega^{kl}+\frac{\Lambda }{6} \vartheta^{i}\wedge \vartheta^{j}\wedge \vartheta^{k}\wedge \vartheta^{l}\Big) \eta_{ijkl}, 
\end{align}
where $\Omega$ is the curvature of $\omega $ (spin connection) and $\kappa =8\pi G/c^{4}$. All indices take values $(0, 1, 2, 3)$. $\mathcal{D}$ means an established 4-dimensional compact domain in spacetime.
In the above formula $\vartheta^{a}$ denote $1$-forms of the Lorentzian coreper in term of which the spacetime looks locally Minkowskian: $g=\eta_{ik} \vartheta^{i}\otimes \vartheta^{k}$, $\eta_{ik}=diag(1,-1,-1,-1)$.\\
$\eta_{ijkl}$ is completely antysymmetric Levi-Civita pseudotensor: $\eta_{0123}=\sqrt{|g|}$, where $g:=det(g_{ik})$. In a Lorentzian coreper $|g|=1$. Spin connection $\omega $ is a general metric connection (or Levi-Civita connection) in Lorentzian coreper. Strictly speaking, by spin connection relativists understand the mixed components $\omega ^{a}_{\ b\mu }$ of the connection $\omega $.
These mixed components are defined in the following way. Firstly, we decompose connection form $\omega ^{i}_{\ k}$ in terms of the Lorentzian 1-forms $\vartheta ^{l}$
$$\omega ^{i}_{\ k}=\omega ^{i}_{\ kl}\vartheta ^{l}.$$
Then, the decomposition of the forms $\vartheta ^{l}$ in the natural cobasis $dx^{\mu }$
$$\vartheta ^{l}=h^{l}_{\ \mu }dx^{\mu }$$
leads to 
$$\omega ^{i}_{\ k}=\omega ^{i}_{\ kl}h^{l}_{\ \mu }dx^{\mu }=: \omega ^{i}_{\ k\mu }dx^{\mu }.$$
Relativists called the mixed components $\omega ^{i}_{\ k\mu }$ the spin connection.
 
For the geometrical units $G=c=1$ the formula (1) takes the form in terms if $\epsilon $ and $\Lambda >0$
\begin{align}
S_{EP}=\frac{1}{32\pi}\int\limits_{\mathcal{D}}\Bigl (\eta _{ijkl}\vartheta^{i}\wedge \vartheta^{j}\wedge \Omega^{kl}+ \frac {\Lambda}{6}\eta _{ijkl}\vartheta^{i}\wedge \vartheta^{j}\wedge \vartheta^{k}\wedge \vartheta^{l}\Bigr). 
\end{align}

Adding to the geometric part $S_{EP}$ the matter action 
\begin{align}
S_{m}=\int \limits_{\mathcal{D}}L_{mat}(\phi ^{A}, D\phi ^{A}, \vartheta^{i}),
\end{align}
where $\phi ^{A}$ means tensor-valued matter form and $D\phi ^{A}$ its absolute exterior derivative, we obtain full action
\begin{align}
S&=S_{EP}+S_{m} \notag
\\&=\frac{1}{32\pi}\int\limits_{\mathcal{D}}\Bigl (\eta _{ijkl}\vartheta^{i}\wedge \vartheta^{j}\wedge \Omega^{kl}+ \frac {\Lambda}{6}\eta _{ijkl}\vartheta^{i}\wedge \vartheta^{j}\wedge \vartheta^{k}\wedge \vartheta^{l}\Bigr)+\int \limits_{\mathcal{D}}L_{mat}(\phi ^{A}, D\phi ^{A}, \vartheta^{i})
\end{align}

After some calculations one gets that the variation $\delta S=\delta S_{EP}+\delta S_{m}$ with respect to $\vartheta^{i},\ \omega^{i}_{\ j}$ and $\phi^{A}$ reads

\begin{align}
\delta S&=\int \limits_{\mathcal{D}}\Bigl [ \frac{1}{8\pi }\delta \vartheta^{i}\wedge \Bigl ( \frac{1}{2}\Omega ^{kl}\wedge \eta_{kli}+\Lambda \eta_{i}+8\pi t_{i}\Bigr ) \notag
\\&+\frac{1}{2}\delta \omega ^{i}_{\ j}\wedge\Bigl (\frac{1}{8\pi }D\eta^{\ j}_{i}+s_{i}^{\ j}\Bigr )+\delta \phi^{A}\wedge L_{A}+ an \ exact \ form\Bigr ]. 
\end{align}

The three-forms: energy-momentum $t_{i}$, classical spin $s_{i}^{\ j}$ and $L^{A}$ are defined by the following form of the variation $\delta L_{m}$
\begin{align}
\delta L_{m}=\delta \vartheta^{i}\wedge t_{i}+\frac{1}{2}\delta \omega^{i}_{\ j}\wedge s^{\ j}_{i}+\delta \phi ^{A}\wedge L^{A}+ an \ exact \ form.
\end{align}
$\eta_{kli},\ \eta_{i}^{\ j}, \eta_{i}$ mean the forms introduced in the past by A. Trautman [11]. 

The variations $\delta\vartheta^{i}$, $\delta \omega^{i}_{\ j}$ and $\delta \phi^{A}$ are vanishing on the boundary $\partial \mathcal{D}$ of the compact domain $\mathcal{D}$.

Einstein's equations like all the other physical field equations can be obtained from the variational principle, which is called the Principle of Stationary Action or Hamiltonian Principle. In our case it has the following form:
\begin{align}
\delta S=0, 
\end{align}

It leads us to the following sets of the field equations
\begin{align}
 \frac{1}{2}\Omega ^{kl}\wedge \eta_{kli}+\Lambda \eta_{i}=-8\pi t_{i}
\end{align}
\begin{align}
D\eta^{\ j}_{i}=\eta ^{\ j}_{i \ \ k}\wedge \Theta ^{k}=-8\pi s_{i}^{\ j} 
\end{align}
and
\begin{align}
L_{A}=0. 
\end{align}
$L_{A}=0$ represent equations of motion for matter field. These equations are not intrinsic in further our considerations, so we will omit them. We are interested only in the gravitational field equations which are given by the equations (8)-(9).
The equations (8)-(10) (but with $\lambda $=0) were obtained at first by A. Trautman [11]. Only one difference is in our approach - the opposite sign of the spin tensor.

In vacuum where $t_{i}=s^{\ j}_{i}=0 \implies \eta^{\ j}_{i\ \ k}\wedge \Theta ^{k}=0 \implies \Theta ^{k}=0$, i.e. vanishing torsion and we get the standard vacuum Einstein's equations (EE) with cosmological constant $\Lambda$ and pseudoriemannian geometry 
\begin{align}
 \frac{1}{2}\Omega ^{kl}\wedge \eta_{kli}+\Lambda \eta_{i}=0. 
\end{align}
In general, we have the Einstein-Cartan equations and Riemann-Cartan geometry (a metric geometry with torsion).\\
The standard GR we obtain also if we put $\frac{\delta L_{m}}{\delta \omega ^{i}_{\ k}}=0 \implies s_{i}^{\ k}=0 \implies \eta_{i\ \ j}^{\ k}\wedge \Theta ^{j}=0 \implies \Theta ^{j}=0$. It is GR inside spinless matter with equations
\begin{align}
\frac{1}{2}\Omega^{kl}\wedge \eta_{kli}+\Lambda \eta_{i}=-8\pi t_{i}. 
\end{align}

One can show that $\frac{1}{2}\Omega^{kl}\wedge \eta_{kli}=-G_{i}^{s}\eta_{s}$ and $t_{i}=T_{i}^{\ s}\eta_{s}$, where the Einstein tensor $G_{i}^{s}$ is defined as follows
\begin{align}
G_{i}^{s}=R_{i}^{s}-\frac{1}{2}\delta_{i}^{s}R.
\end{align}
Putting $t_{i}=T^{\ s}_{i}\eta_{s}$ we get from (12)
\begin{align}
-G_{i}^{\ s}\eta_{s}+\Lambda \delta_{i}^{s}\eta _{s}=-8\pi T_{i}^{\ s}\eta_{s}. 
\end{align}
or
\begin{align}
G_{i}^{\ s}-\Lambda \delta_{i}^{s}=8\pi T_{i}^{\ s}.
\end{align}
(15) are standard Einstein equations with cosmological constant $\Lambda $ in tensorial notation with symmetric matter tensor: $T^{ik}=T^{ki}$.

\section{Results: Einstein-Palatini action integral for General Relativity in vacuum and with positive cosmological constant $\Lambda $ as integral action for a gauge field}

Let us get back to Einstein-Palatini action (1) in vacuum and let introduce the duality operator $\star $ [1]
\begin{align}
\star :=-\frac{\eta_{ijkl}}{2} \quad \implies \eta_{ijkl}=-2\star .
\end{align}
Then one has
\begin{align}
\eta_{ijkl}\Omega^{kl}=-2\star \Omega_{ij},
\\ \eta_{ijkl}\vartheta^{k}\wedge \vartheta^{l}=-2\star \bigl (\vartheta_{i}\wedge \vartheta_{j}\bigr )
\end{align}
and the Einstein-Palatini action has the following form
\begin{align}
S_{EP}&=-\frac{1}{2\kappa}\int \limits_{\mathcal{D}}\Bigl (\vartheta^{i}\wedge \vartheta^{j}\wedge \star \Omega_{ij}+\frac{\Lambda}{6} \vartheta^{i}\wedge \vartheta^{j}\wedge \star \bigl (\vartheta_{i}\wedge \vartheta_{j}\bigr )\Bigr )\notag\\
&=-\frac{1}{2\kappa}\int \limits_{\mathcal{D}}tr\Bigl (\vartheta \wedge \vartheta \wedge \star \Omega +\frac{\Lambda}{6} \vartheta \wedge \vartheta \wedge \star\bigl (\vartheta \wedge \vartheta \bigr )\Bigr ).
\end{align}
Now, let us introduce the corrected curvature $\Omega_{cor}$
\begin{align}
\Omega_{cor}:=\Omega+\frac{\Lambda}{3}\vartheta \wedge \vartheta \quad \implies \vartheta \wedge \vartheta =-\frac{3}{\Lambda}\Bigl (\Omega-\Omega_{cor}\Bigr ). 
\end{align}
Substituting the last formula into Einstein-Palatini action we get 
\begin{align}
S_{EP}&=\frac{1}{2\kappa}\int \limits_{\mathcal{D}}tr\Bigl (\vartheta \wedge \vartheta \wedge \star \Omega+\frac{\Lambda}{6} \vartheta \wedge \vartheta \wedge \star\bigl (\vartheta \wedge \vartheta \bigr )\Bigr )\notag \\
&=\frac{1}{2\kappa}\int \limits_{\mathcal{D}} tr\Bigl [\frac{3}{\Lambda}\bigl (\Omega-\Omega_{cor}\bigr )\wedge\star \Omega -\frac{\Lambda}{6}\frac{9}{\Lambda^{2}}\bigl (\Omega-\Omega_{cor}\bigr )\wedge\star \bigl (\Omega -\Omega_{cor}\bigr )\Bigr ] \notag \\
&=\frac{3}{4\Lambda \kappa}\int \limits_{\mathcal{D}} tr\Bigl (2\bigl (\Omega-\Omega_{cor}\bigr )\wedge \star \Omega-\bigl (\Omega -\Omega_{cor}\bigr )\wedge \star \bigl (\Omega -\Omega_{cor}\bigr )\Bigr ) \notag \\
&=\frac{3}{4\Lambda \kappa}\int \limits_{\mathcal{D}} tr\Bigl [2\Omega \wedge \star \Omega -2\Omega_{cor}\wedge \star \Omega -\Omega \wedge \star \Omega +\Omega_{cor}\wedge \star \Omega +\Omega \wedge \star \Omega_{cor}-\Omega_{cor}\wedge \star \Omega_{cor}\Bigr ] \notag \\
&=\frac{3}{4\Lambda \kappa}\int \limits_{\mathcal{D}} tr\Bigl [\Omega \wedge \star \Omega -\Omega_{cor}\wedge \star \Omega +\Omega \wedge \star \Omega_{Cor}-\Omega_{cor}\wedge \star \Omega_{cor}\Bigr ] 
\end{align}
Because $-\Omega_{cor}\wedge \star \Omega +\Omega \wedge \star \Omega_{cor}$ reduces, then we finally have
\begin{align}
S_{EP}=\frac{3}{4\Lambda \kappa}\int \limits_{\mathcal{D}} tr\Bigl [\Omega \wedge \star \Omega -\Omega_{cor}\wedge \star \Omega_{cor}\Bigr ].
\end{align}

The expression $tr\bigl (\Omega \wedge \star \Omega \bigr )=\eta_{ijkl} \Omega ^{ij}\wedge \Omega ^{kl}$ is in four dimensions a topological invariant called Euler's form, which does not influence the equations of motion [12]. Hence, in 4-dimensions the Einstein-Palatini action is equivalent to 

\begin{align}
S_{EP}=-\frac{3}{4\Lambda \kappa}\int \limits_{\mathcal{D}} tr\Bigl (\Omega_{cor}\wedge \star\Omega_{cor}\Bigr ). 
\end{align}

We see that the Einstein-Palatini action in 4-dimensions is efectively the functional which is quadratic function of the corrected Riemannian curvature, i.e., it has a form of the action for a gauge field.

The only difference is that in (23) we have the star operator $\star $, which is different from Hodge star operator. Namely, our star operator acts onto "interior" indices (tetrad's indices), not onto forms as Hode duality operator does.

The gauge group for the theory with action (23) is the Lorentz group $\mathcal {L}=SO(1,3)$ or its double cover $SL(2,\mathbb{C})$.

It is interesting that $\Omega_{cor}=0$ for the de Sitter spacetime which is the fundamental vacuum solution to the Einstein equations
\begin{align}
G_{i}^{\ s}-\Lambda \delta_{i}^{s}=0.
\end{align}
We would like to emphasize that in the case $\Lambda=0$ the above trick with $\Omega_{cor}$ breaks. Namely, we have in this case (see Section 3) $\Omega_{cor}=\Omega $ because $[e_{i},e_{k}]=0$. This result formally trivializes $S_{E-P}$ action to the strange form $S_{E-P}=0$. It is easily seen from (21) or (22). In the case $\Lambda<0$ one obtains the result analogical to (23) with $\Omega_{cor}=\Omega +\frac{\Lambda}{3}\vartheta \wedge \vartheta$ but this time $\Lambda <0$. We did not consider this case because it needs to introduce into calculations the anti-de Sitter spacetime (and its isometry group SO(2,3)) which has very strange casual properties.

\newpage
\section{Discussion: Geometrical interpretation of the corrected curvature $\Omega_{cor}$}

Let $P(M_{4}, GdS)$ denote the principal bundle of de Sitter basis over a manifold $M_{4}$ (space-time) with de Sitter group($ GdS$) [5, 13] as a structure group. This group is isomorphic to the group $SO(1, 4)$ [3, 5, 13]. Let $\widetilde{\omega}$ be 1-form of connection in the principle fibre  bundle $P(M_{4}, GdS)$. The form $\widetilde{\omega }$ has values in the algebra $\mathfrak{g}$ of the group $GdS$. This algebra splits (as a vector space) into direct sum
\begin{align}
\mathfrak{g}=so(1, 3)\oplus R^{(1, 3)}.
\end{align}
$so(3, 1)$ denotes here algebra of the group SO(1, 3), which is isomorphic to Lorentz group $\mathcal {L}$, and $R^{(1, 3)}$ is a 4-dimensional vector space of generalised translations (translations in the curved de Sitter spacetime). One can identify the de Sitter spacetime with the quotient SO(1,4)/SO(1,3). \\
Let us define $so(1,3)=:\mathfrak{h}$, $R^{1,3}=:\mathfrak{p}$. Then we have [1,2]
\begin{align}
\mathfrak{g}=\mathfrak{h}\oplus \mathfrak{p},
\end{align}
and 
\begin{align}
[\mathfrak{h},\mathfrak{h}]\subset \mathfrak{h},\ [\mathfrak{h},\mathfrak{p}]\subset \mathfrak{p},\ [\mathfrak{p},\mathfrak{p}]\subset \mathfrak{h}.
\end{align}
This means that the Lie algebra $\mathfrak{g}$ is a symmetric Lie algebra [1,2].\\
On the other hand, the spaces which satisfy (26)-(27) are called globally symmetric Riemannian spaces [13].

Let $P(M_{4}, \mathcal {L})$  denote the principal bundle of Lorentz basis over the manifold $M_{4}$. There exists a morphism of principal bundles 
\begin{align}
f:P(M_{4}, \mathcal {L}) \longrightarrow P(M_{4}, GdS)
\end{align}
analogical to the morphism of the bundle linear frames and the bundle affine frames [4].
This morphism creates pull-back $f_{*}\widetilde{\omega }$ of the form $\widetilde{\omega}$ onto the bundle $P(M_{4}, \mathcal {L})$. Here $\widetilde{\omega}$ is the connection 1-form in the bundle $P(M_{4},GdS)$. \\
Let us denote this pull-back by $A$. $A$ is a 1-form on $P(M_{4}, \mathcal {L})$ with values in the direct sum [4]
\begin{align}
so(1, 3)\oplus R^{(1, 3)}.
\end{align}

Hence, we have a natural decomposition [4]
\begin{align}
A=f_{*}\widetilde{\omega}=\omega+\theta ,
\end{align}
where $\omega $ is a 1-form on $P(M_{4}, \mathcal {L})$ with values in the algebra $so(1, 3)$ and $\theta $ is a 1-form on $P(M_{4}, \mathcal {L})$ with values on $R^{(1, 3)}$. $\omega $ is a connection on the bundle $P(M_{4}, \mathcal {L})$.

On the base $M_{4}$ the 1-form $\theta $ can be identified with 1-form $\vartheta $ already used in this paper: $\theta=\vartheta $. In the following we will work on the base space $M_{4}$ and write (30) in the form 
$$A=\omega + \vartheta .$$

Let us compute a 2-form curvature $\widetilde{\Omega}$ of the pulled back $A$. From the definition we have
\begin{align}
\widetilde{\Omega}&=dA+\frac{1}{2}\Bigl [A,A\Bigr ] \notag \\
&=d\bigl (\omega+\vartheta\bigr )+\frac{1}{2}\Bigl [\omega+\vartheta,\omega+\vartheta\Bigr ] \notag \\
&=d\omega+\frac{1}{2}\Bigl [\omega,\omega\Bigr ]+d\vartheta+\frac{1}{2}\Bigl [\omega, \vartheta\Bigr ]+\frac{1}{2}\Bigl [\vartheta,\omega\Bigr ]+\frac{1}{2}\Bigl [\vartheta,\vartheta\Bigr ].  
\end{align}

We are going to introduce to our equations bases $\widetilde{M}_{ik}=-\widetilde{M}_{ki}$ of algebra $so(1, 3)$ and ${e_{i}}$ of vector space $R^{(1, 3)}$. In these bases, we have 
\begin{align}
\omega =\omega ^{i}_{\ k}\widetilde{M}^{\ k}_{i}=\omega ^{ik}\widetilde{M}_{ik},\  \vartheta =\vartheta ^{i}e_{i}.
\end{align}

$(\widetilde{M}_{ik},e_{l})$ form together the algebra of the de Sitter group (the basis of algebra $\mathfrak {g}$). Our elements $\widetilde{M}_{ik}=-\widetilde{M}_{ki}$ are real and connected with elements $M_{ik}=M_{ki}$ used in [13] in the following way 
\begin{align}
\widetilde{M}_{ik}=\frac{i}{2}M_{ik} \Rightarrow M_{ik}=-2i\widetilde{M}_{ik}.
\end{align} 
The comutational relations for the algebra $so(1, 4)=so(1, 3)\oplus R^{(1, 3)}$, in the terms of the elements $[\widetilde{M}_{ik}, e_{l}]$, read
\begin{align}
&[\widetilde{M}_{ij}, \widetilde{M}_{kl}]=\frac{1}{2}\bigl (\eta _{il}\widetilde{M}_{jk}+\eta _{jk}\widetilde{M}_{il}-\eta _{ik}\widetilde{M}_{jl}-\eta _{jl}\widetilde{M}_{ik}\bigr )  
\\&[e_{i}, \widetilde{M}_{jk}]=\frac{1}{2}\bigl (\eta _{ij}e_{k}-\eta _{ik}e_{j}\bigr )  
\\&[e_{i}, e_{j}]=\frac{2\widetilde{M}_{ij}}{R^{2}}  
\end{align}
The following commutation relations are important in the further considerations [5, 6, 13].
\begin{align}
&[\widetilde{M}_{ki}, e_{l}]=\frac{1}{2}\bigl (\eta_{il}e_{k}-\eta_{kl}e_{i}\bigr ) 
\\ &[e_{i}, e_{k}]=\frac{2 \widetilde{M}_{ik}}{R^{2}}, 
\end{align}
where $R$ is the radius of the de Sitter spacetime. This radius $R$ is connected with $\Lambda $ by the formula $\Lambda=\frac{3}{R^{2}}$. Using the above equations we have
\begin{align}
\widetilde {\Omega}=\Omega_{\omega }+\frac{1}{2}\omega^{i}_{\ k}\wedge\vartheta^{l}\Bigl [\widetilde{M}^{k}_{\ i},e_{l}\Bigr ]+\frac{1}{2}\vartheta^{l}\wedge\omega^{i}_{\ k}\Bigl [e_{l},\widetilde{M}^{k}_{\ i}\Bigr ]+\frac{1}{2}\vartheta^{i}\wedge\vartheta^{k}\Bigl [e_{i},e_{k}\Bigr ]+d\vartheta^{i}e_{i}, 
\end{align}
where $\Omega _{\omega}=d\omega +\frac{1}{2}\bigl [\omega, \omega \bigr ]$ is the curvature 2-form of the connection's $\omega $.
Taking into consideration the commutation relations in algebra $\mathfrak{g}$ given by the formulas (37) and (38), we obtain
\begin{align}
\widetilde {\Omega}&=\Omega_{\omega }+\omega^{i}_{\ k}\wedge\vartheta^{l}\Bigl [\widetilde{M}^{\ k}_{i},e_{l}\Bigr ]+\vartheta^{i}\wedge\vartheta^{k}\frac{\widetilde{M}_{ik}}{R^{2}}+d\vartheta^{i}e_{i} \notag \\
&=\Omega^{ik}_{\omega }\widetilde{M}_{ik}+\bigl (\omega^{i}_{.k}\wedge\vartheta^{k}\bigr )e_{i}+d\bigl (\vartheta^{i}\bigr )e_{i}+\frac{\vartheta^{i}\wedge\vartheta^{k}\widetilde{M}_{ik}}{R^{2}}\notag \\
&=\Omega^{ik}_{\omega }\widetilde{M}_{ik}+\frac{\vartheta^{i}\wedge\vartheta^{k}\widetilde{M}_{ik}}{R^{2}}+\bigl (d\vartheta^{i}+\omega^{i}_{.k}\wedge\vartheta^{k}\bigr )e_{i} \notag \\
&=\Omega ^{ik}_{corr}\widetilde{M}_{ik}+\bigl (\mathcal{D_{\omega}}\vartheta^{i}\bigr )e_{i} \notag \\
&=\Omega ^{ik}_{corr}\widetilde{M}_{ik}+\Theta ^{i}e_{i}
\end{align}
$\Omega _{cor}:=\Omega _{\omega }+\frac{\vartheta \wedge \vartheta }{R^{2}}=\Omega _{\omega }+\frac{\Lambda }{3}\vartheta \wedge \vartheta $ and it denotes the corrected curvature of the connection $\omega $ on the bundle $P(M_{4}, \mathcal L)$ and $\Theta =\mathcal D_{\omega }\vartheta $ is a torsion of the connection $\omega $.
\\ If we adjust the connection $\widetilde{\omega }$ in such a way that the connection $\omega $ is torsionless ($\Theta =0$), i.e. if $\omega$ is Levi-Civita connection, then we get (after leaving the basis $so(1, 3)$ and $R^{(1, 3)}$) 
\begin{align}
\widetilde {\Omega}=\Omega _{\omega }+\frac{\Lambda }{3}\vartheta \wedge \vartheta =\Omega _{cor}\ .
\end{align}

In the Section 2 we gave the definition of the corrected curvature $\Omega_{cor}$ as follows:
\begin{align}
\Omega_{cor}:=\Omega+\frac{\Lambda}{3}\vartheta \wedge \vartheta. 
\end{align}
As one can see this curvature is a curvature of the connection
\begin{align}
A:=f _{*}\widetilde{\omega }=\omega +\vartheta 
\end{align}
if $\Theta =0,$ e.g., in Einstein-Cartan vacuum.\\
If $\Theta \neq 0$, e.g. in Einstein theory with spinning sources, then $\Omega_{cor}$ is the $so(1,3)$-part of the curvature $\widetilde {\Omega}$.

\section{Conclusion}

In this article we have shown that in four dimensions the action integral for GR with a positive cosmological constant $\Lambda $ can be written in an analogical form to the form of the action integral for the typical gauge field. However, there is one difference - the star. Instead of the Hodge star, we have slightly different star called the duality operator [2, 12].\\
Our result is important because it shows that there is no need to generalize GR and construct very complicated gravitational theories to obtain a gravitational theory as a gauge theory. The ordinary GR formulated in terms of tetrads and spin connection with cosmological constant $\Lambda $ > $0$ is already a gauge theory with gauge group $\mathcal {L}=SO(1,3)$ or its double cover $SL(2,\mathbb {C})$. This fact is very interesting in connection with universality of the Einstein theory: every alternative metric theory of gravity can be reformulated as Einstein theory with additional "egzotic" matter fields [15,16]. Therefore we present a following conjecture.\\
\\
\\
Conjecture: 

After the above reformulation one can put the pure geometric part of the action (identic with the geometric part $S_{EP}$) for any alternative theory with $\Lambda >0$ in the form (23). This Conjecture will be studied in future.\\
Some scientists [1, 2, 3] were concerned with this problem and they came to the similar conclusions as ours, but they applied in their works the Cartan's approach to the connection in the principal bundle [2, 13, 14]. This approach is not well known among geometrists and relativists. We have used only the standard theory of connection in the principal bundle which was created by Ehresmann - Cartan's student [4, 8]. His approach is commonly used in differential geometry and in relativity. We would like to emphasize that the formulation of the EP action in the form (23) can be important for quantizing of general relativity (because gauge fields can be succesfully quantized).

\newpage
\section*{Appendix}
\begin{center}
\underline {$\eta$ forms and operations with them [11]}
\end{center}

Following [11] we define
\begin{align}
\eta _{ijkl}=\sqrt{|g|}\epsilon _{ijkl} \tag {A.1}
\end{align}
where $\epsilon _{ijkl}$ is Levi-Civita pseudotensor with properties
\begin{align}
\epsilon _{ijkl}=\begin{cases}\   \ 1 &\text {if the sequence of indices ijkl is an even permutation}\\ & \text{of the sequence 0, 1, 2, 3}; \\-1 &\text{if it is an odd permutation}; \\ \ \  0 &\text{if the sequence of indices ijkl is not an even permutation}\\ &\text{of the sequence 0, 1, 2, 3} \end{cases}.\tag {A.2}
\end{align}
and we take $\eta_{0123}=\sqrt{|g|}$. In Lorentzian coreper $|g|=1$.

One has [11]
\begin{align}
&\eta_{ijk}=\vartheta ^{l}\wedge \eta _{ijkl} \tag {A.3}
\\&\eta _{ij}=\frac{1}{2}\vartheta ^{k}\wedge \eta _{ijk} \tag {A.4}
\\&\eta _{i}=\frac{1}{3}\vartheta ^{j}\wedge \eta _{ij} \tag {A.5}
\\&\eta =\frac{1}{4}\vartheta ^{i}\wedge \eta _{i} \tag {A.6}
\\&\vartheta ^{n} \eta _{ijkl}=\delta ^{n}_{l}\eta _{ijn}+\delta ^{n}_{j}\eta _{lik}-\delta ^{n}_{i}\eta _{jkl}-\delta ^{n}_{k}\eta _{lij} \tag {A.7}
\\&\vartheta ^{n}\wedge \eta _{kli}=\delta ^{n}_{i}\eta _{kl}+\delta ^{n}_{l}\eta _{ik}+\delta ^{n}_{k}\eta _{li} \tag {A.8}
\\&\vartheta ^{m}\wedge \eta _{kl}=\delta ^{m}_{l}\eta _{k}-\delta ^{m}_{k}\eta _{l} \tag {A.9}
\\&\vartheta ^{j}\wedge \eta _{i}=\delta ^{j}_{i}\eta \tag {A.10}
\end{align}
The forms $\eta $,  $\eta_{i}$,  $\eta_{ij}$,  $\eta_{ijk}$ are Hodge dual to the forms $1$, $\vartheta^{i}$, $\vartheta^{i}\wedge \vartheta^{j}$, $\vartheta^{i}\wedge \vartheta^{j}\wedge \vartheta^{k}$ respectively [11].

\end{document}